\documentclass[aps,  amsmath,  amssymb,  amsfonts,  showpace,  twocolumn,  pra]{revtex4}

\usepackage{diagbox}
\usepackage{amscd}
\usepackage{amsthm}
\usepackage{graphicx}
\usepackage{epstopdf}
\usepackage{mathdots}
\usepackage{blkarray}
\makeatletter
\newbox\BA@first@box
\makeatother

\newtheorem{Thm}{Theorem}
\newtheorem{Lem}{Lemma}
\newtheorem{Cor}{Corollary}
\newtheorem{Prop}{Proposition}
\newtheorem{Exm}{Example}

\usepackage[
  colorlinks,
  linkcolor = blue,
  citecolor = blue,
  urlcolor = blue]{hyperref}

\usepackage{amsmath,  amssymb,  amsthm}

\usepackage{tikz}
\usetikzlibrary{arrows}
\usetikzlibrary{calc,  positioning}

\usepackage{thm-restate}

\usepackage{colonequals}

\usepackage[font=small]{caption}

\newtheoremstyle{noCaption}
{\topsep}
{\topsep}
{\itshape}
{}
{}
{}
{0pt}
{}%

\begin{document}
\title{Genuinely nonlocal sets with smallest cardinality}
\author{Zong-Xing Xiong$^{1}$, Mao-Sheng Li$^{2}$, Bing Yu$^{3}$, Zhu-Jun Zheng$^{2}$, Lvzhou Li$^{1}$\footnote{Contact author: lilvzh@mail.sysu.edu.cn}\\
{\footnotesize{\it $^1$ Institute of Quantum Computing and Software, School of Computer Science and Engineering, Sun Yat-Sen University, Guangzhou 510006, China}}\\
{\footnotesize{\it $^2$ School of Mathematics, South China University of Technology, Guangzhou 510641, China}}\\
{\footnotesize{\it $^3$ School of Mathematics and Systems Science, Guangdong Polytechnic Normal University, Guangzhou 510665, China}}\\
}

\affiliation{
}

\begin{abstract}
Recently, there is growing interest in the study of genuine nonlocality, which serves to explore the local accessability of global information encoded in orthogonal multipartite quantum states under scenarios where not all subsystems are joined together. For such form of nonlocality, a probably most fundamental question is upon what states it is prone to be manifested. To tackle this, we present in this work genuinely nonlocal sets with the smallest possible cardinality. We first show the existence of genuinely nonlocal sets of three pure states in arbitrary $N$-partite system. As a byproduct, this also gives new examples of strongly nonlocal sets with dramatically smaller cardinality than ever for all possible systems, settling some related questions effortlessly. Then, for mixed hypothetical states, we show that genuinely nonlocal sets of two even exist, regardless of the number of copies available. In particular, it turns out for both our constructions that certain genuinely entangled states necessarily exist, nontrivially indicating their potential of raising difficulty in locally accessing multipartite quantum information.
\end{abstract}
\pacs{03.  67.  Hk,  03.  65.  Ud }
\maketitle

\section{Introduction} \label{section1}
In 1999, C.H.Bennett et al. introduced a new form of quantum nonlocality, which serves to investigate the fundamental question about locally accessing quantum information encoded in a set of quantum states \cite{Bennett99}. In his seminal work, Bennett (and his co-workers) showed the existence of nine product states in bipartite system $\mathbb{C}^3 \otimes \mathbb{C}^3$ that cannot be perfectly distinguished by the separated subsystems, provided only local operations and classical communications (LOCC) are available by them. He named this phenomenon ``quantum nonlocality without entanglement'', as opposed to the stereotype of many people that to manifest nonlocality, existence of entanglement is always required. Since then, such form of distinguishability-based nonlocality has been studied extensively (see \cite{Bennett1999,Walgate00,Walgate02,Ghosh01,Ghosh02,DiVicenzo03,Horodecki03,Chefles04,Hayashi06,Nathanson05,Bandyopadhyay09,Bandyopadhyay11,Li2020,Xu1,Xu2} for an incomplete list).

Although entanglement is not necessary for this form of nonlocality, lots of evidence have been found showing that entangled states are generally more difficult to be distinguished locally. For example, it has been shown by Horodecki et al. that if a set of full orthogonal basis $\{|\psi_1\rangle, \cdots, |\psi_{mn}\rangle\}$ of any bipartite system $\mathbb{C}^m \otimes \mathbb{C}^n$ is locally distinguishable, then they must be all product states \cite{Horodecki03}. Moreover, if at least one among them is entangled, then the states are not even conclusively distinguishable locally. In \cite{Hayashi06}, it was further shown by Hayashi et al. that the number $k$ of orthogonal pure states in $\mathbb{C}^m \otimes \mathbb{C}^n$ that is perfectly distinguishable by LOCC is bounded above by the total dimension of the two-partite system over the average entanglement of the states, namely $k \leq mn/[1+\overline{r(\psi_i)}]$ where $r(\cdot)$ is the robustness of entanglement. With that said, entanglement is generally responsible for local indistinguishability, though there also exists certain exceptional occasions \cite{Horodecki03}.

Notably, it can be deduced from the result of Hayashi et al. that any $d+1$ orthogonal maximally entangled states in bipartite system $\mathbb{C}^d \otimes \mathbb{C}^d$ must be locally indistinguishable. This fact was first observed by Nathanson \cite{Nathanson05} , who also showed that any three maximally entangled states in $\mathbb{C}^3 \otimes \mathbb{C}^3$ are locally distinguishable. For cases $d \geq 4$, examples were also found by researchers that $k \leq d$ locally indistinguishable orthogonal maximally entangled states exist in $\mathbb{C}^d \otimes \mathbb{C}^d$ \cite{Yu12,Cosentino13,CosentinoR14,Li15,Yu15}. In particular, Yu and Oh showed that the the minimal number of locally indistinguishable maximally entangled states in $\mathbb{C}^d \otimes \mathbb{C}^d$ can scale asymptotically down to $3d/4$ when $d$ is large. Nevertheless, a seemingly most simple problem whether there exist three locally indistinguishable maximally entangled states in $\mathbb{C}^d \otimes \mathbb{C}^d$ for $d \geq 4$ remains unsolved today. What is known is that any two orthongonal quantum states (whether entangled or not) can always be locally distinguished, a well-known result established by Walgate \cite{Walgate00}.

Generally speaking, a set of more quantum states might usually tend to be harder for locally distinguishing, while a set with less states is often more likely to be distinguishable. So if one would like to find small sets of quantum states that manifest nonlocality, some nontrivial features ought to be possessed by the sets. In \cite{Bennett1999}, Bennett et. al. discovered the existence of another set of five locally indistinguishable product states in $\mathbb{C}^3 \otimes \mathbb{C}^3$, namely the ``unextendible product basis'' (UPB). It was also proved latter that any four or less orthogonal product states in any bipartite system in locally distinguishable \cite{DiVicenzo03}. Arguably, this can be regarded as yet another evidence showing that entanglement can raise difficulty in local discrimination of orthogonal quantum states, for locally indistinguishable sets of three or four exist when entangled states are present. For example, the three Bell states
\begin{equation}\begin{split}\label{Bell}
& |\beta_{1}\rangle = \frac{1}{\sqrt{2}} (|00\rangle + |11\rangle), \ \ |\beta_{2}\rangle = \frac{1}{\sqrt{2}} (|00\rangle - |11\rangle), \\
& |\beta_{3}\rangle = \frac{1}{\sqrt{2}} (|01\rangle + |10\rangle)
\end{split}\end{equation}
in $\mathbb{C}^2 \otimes \mathbb{C}^2$ was first observed by Ghosh et al. to be locally indistinguishable \cite{Ghosh01}. For one more evidence, it has also been proved that all sets of orthogonal product states in systems $\mathbb{C}^2 \otimes \mathbb{C}^d$ $(d\geq 2)$ are locally distinguishable \cite{DiVicenzo03}.

Moving to the scenarios where $N > 2$ subsystems are considered, we should notice that Walgate's result about local distinguishability of two orthogonal states also holds when the subsystems are separated from each other and only local measurements are allowed. Then a question naturally raise: what is the smallest set of orthogonal states that cannot be locally distinguished for $N$-partite systems ($N > 2$)? The answer to this problem is however trivial: one can easily construct locally indistinguishable sets of three multipartite states directly from the bipartite ones, say those states in (\ref{Bell}). For example, given any bipartition $L|R$ of the $N$ subsystems, the states
\begin{equation}\begin{split}\label{2}
& |\beta_{1}\rangle_{lr} \ \otimes \ |0\rangle^{\otimes N-2}, \\
& |\beta_{2}\rangle_{lr} \ \otimes \ |0\rangle^{\otimes N-2}, \\
& |\beta_{3}\rangle_{lr} \ \otimes \ |0\rangle^{\otimes N-2} \ \ \ (l \in L, r \in R)
\end{split}\end{equation}
or alternatively, the states
\begin{equation}\begin{split}\label{3}
&  \frac{1}{\sqrt{2}} \left(|0\rangle^{\otimes N} + |1\rangle^{\otimes N}\right), \\
&  \frac{1}{\sqrt{2}} \left(|0\rangle^{\otimes N} - |1\rangle^{\otimes N}\right), \\
&  \frac{1}{\sqrt{2}} \left(|0\rangle^{\otimes |L|}|1\rangle^{\otimes |R|} + |1\rangle^{\otimes |L|}|0\rangle^{\otimes |R|}\right)
\end{split}\end{equation}
are locally indistinguishable, for both sets are locally equivalent to (\ref{Bell}) across the bipartition $L|R$ and indistinguishability in such a weak separation does certainly implies indistinguishability when all subsystems are fully separated. Nevertheless, in recent literature, there is another naturally arising form of nonlocality that has been considered intensively, namely, local indistinguishability of the states in situations where not all subsystems are joined together \cite{Rout19, Halder19, Li21,Rout21, shi, xiong2023, xiong2024, Lu24, Wang21, Shi22, Zhou23, He24, Li23, liji, Hu24, Zhen24}. Such form of nonlocality is termed \emph{genuine nonlocality} where ``genuine'' stands for ``all bipartitions'', in analogy with the well-known definition of genuine multipartite entanglement. Much similar to the bipartite scenario, it's fundamental to ask upon what states such nonlocality is prone to be manifest. Or, what is the smallest set of orthogonal $N$-partite states to manifest genuine nonlocality? Does the existence of genuine multipartite entanglement raise difficulty in locally accessing global information? In this work, we are probing such problems for both the case of pure hypothetical states and the case of mixed hypothetical states.

The article is organized as follows. In Sec. \ref{section2}, we review some of the necessary backgrounds and definitions. In Sec. \ref{section3}, we show that in arbitrary $N$-partite system, there always exists genuinely nonlocal set of three orthogonal pure states, among which two are the most typical genuine entangled states -- the $N$-qubit GHZ pair. As a byproduct, this also gives new examples of strongly nonlocal sets for all $N$-partite systems, with cardinality dramatically smaller than all known ones. In Sec. \ref{section4}, we show that genuinely nonlocal sets of two orthogonal mixed states even exist in all $N$-partite systems. Intriguingly, for such states genuine nonlocality persists regardless of the number of copies available. In Sec. \ref{section5}, we draw our conclusion and discuss some related problems.

\medskip
\section{Preliminaries} \label{section2}
In this section, we provide some relevant backgrounds and definitions that are necessary in this article.

\smallskip
\emph{Local distinguishability (LOCC-distinguishability).} A set of orthogonal hypothetic quantum states $S = \{\rho_1, \rho_2,$ $\cdots, \rho_k\}$ (either pure or mixed) of the multipartite system $\mathcal{H} = \bigotimes_{i=1}^N \mathcal{H}_i$ is called locally distinguishable if the separated subsystem observers are able to tell deterministically which state they share through some protocol of local operations (measurements) and classical communications (LOCC).

\smallskip
\emph{Local irreducibility.} In \cite{Halder19}, Halder et al. put forward the notion of local irreducibility. A set of orthogonal multipartite quantum states is locally irreducible if it is impossible to eleminate one or more states from the whole set with the restriction that only orthogonality--preserving local measurements (OPLMs) are allowed. By definition, a set of locally irreducible multipartite quantum states must be locally indistinguishable, while the opposite is generally not true.

\smallskip
\emph{Genuine nonlocality.} A set of orthogonal multipartite quantum states is called genuinely nonlocal if the states are not locally distinguishable across any bipartition of the subsystems. By definition, genuine nonlocality implies local indistinguishability while the opposite is generally not true. In \cite{Halder19, Rout19, Li21,Rout21, shi, xiong2023, xiong2024, Lu24}, different techniques have been proposed for detecting genuinely nonlocal sets of different types of multipartite quantum states.

\smallskip
\emph{Strong nonlocality.} Among all those techniques for detecting genuinely nonlocal sets of multipartite states, the so called ``trivial orthogonality--preserving local measurement'' (TOPLM), which was originated from Walgate \cite{Walgate02}, is the most frequently exploited one. This technique was first utilized by \cite{Halder19} for constructing genuinely nonlocal sets of product states in several small multipartite systems. In \cite{Halder19}, the authors also raised the definition strong nonlocality, which refers to local irreducibility of the multipartite quantum states across every bipartition of the subsystems. However, it's usually not easy in practice to determine whether a set of states are locally irreducible, except when the subsystems can only perform trivial orthogonality--preserving local measurements. In the literature, though lots of efforts have been paid in seeking strongly nonlocal sets in multipartite systems, all existing examples by far are constructed through the TOPLM technique (see \cite{Wang21,Shi22, Zhou23, He24, Li23, liji, Hu24} for an incomplete list).

\smallskip
\emph{PPT-distinguishability.} A positive semidefinite operator $M \geq 0$ acting on $\mathcal{H}_A \otimes \mathcal{H}_B$ is called a PPT (positive-partial-transpose) operator if its partial transpose about one subsystem (say, $A$) is also positive semidefinite: $M^{T_A} \geq 0$. Note that LOCC measurement operators are separable measurement operators, which are in turn PPT measurement operators, so a set of locally distinguishable bipartite quantum states must also be PPT-distinguishable.

\smallskip
\emph{Conclusively local distinguishability.} A set of orthogonal multipartite quantum states $S = \{\rho_1, \rho_2, \cdots, \rho_k\}$ is conclusively locally distinguishable if every state $\rho_i \in S$ is conclusively locally identifiable, namely, there exists some LOCC protocol whereby with nonzero probability $\rho_i$ can be determined with certainty. By definition, a set of locally distinguishable quantum states must be conclusively locally distinguishable while the opposite is generally not true.

\section{Pure hypothetical states: existence of genuinely nonlocal sets of three}\label{section3}
Since any two orthogonal pure states are locally distinguishable, genuinely nonlocal sets must have cardinality at least three. It's then natural to ask whether genuinely nonlocal sets of three exist in $N$-partite systems. At first sight, this seems somewhat improbable as genuine nonlocality is by definition much stronger than local indistinguishability, especially when a large number of subsystems are considered -- a large proportion $1-\frac1N$ of the subsystems can join to access the information encoded in the set. To manifest genuine nonlocality, it seems necessary that a large number of states are required. Notably, it was recently proved in \cite{Li23} that any genuinely nonlocal set which     is raised with the TOPLM technique in $(\mathbb{C}^d)^{\otimes N}$ must have cardinality at least $d^{N-1}+1$. The work \cite{xiong2024} challenged the effectiveness of this technique by showing the existence of $d+1$ genuinely nonlocal GHZ states in system $(\mathbb{C}^d)^{\otimes N}$. However, to achieve such a cardinality ``$d+1$'', the local dimension $d$ there needs to be sufficiently large as $N$ grows. In \cite{Li21}, genuinely nonlocal sets of product states in $(\mathbb{C}^d)^{\otimes N}$ for $d \geq 3$ was constructed while the cardinality scales linearly in $N$. By all means, it seems unlikely that a constant number of states suffice considering that $N$ can be arbitrarily large and that there are $2^{N-1}-1$ bipartitions being considered.

In what follows, however, it is shown that such seemingly improbable sets exist. More explicitly, we show that genuinely nonlocal sets of three exist in $N$-qubit systems $(\mathbb{C}^2)^{\otimes N}$. Since $(\mathbb{C}^2)^{\otimes N}$ can be embedded into any $N$-product Hilbert space, genuinely nonlocal sets of three also exist in any $N$-partite systems.

To demonstrate, first go back to the bipartite case. Apart from the three Bell states (\ref{Bell}), Ghosh and his co-workers also established the local indistinguishability of the three states
\begin{equation}\begin{split}\label{4}
& |\beta_{1}\rangle = \frac{1}{\sqrt{2}} (|00\rangle + |11\rangle), \ \ |\beta_{2}\rangle = \frac{1}{\sqrt{2}} (|00\rangle - |11\rangle), \\
& |\gamma_{3}\rangle = |01\rangle
\end{split}\end{equation}
in $\mathbb{C}^2 \otimes \mathbb{C}^2$ \cite{Ghosh02}.  They proved this fact by calculating an upper bound on the distillable entanglement of an ingeniously constructed four-party state, \ which will then induce a contradiction provided local distinguishability assumption was made. Actually, there are more concise ways to prove local indistinguishability of these states. For example, it's routine to prove that they are locally irreducible, using the TOPLM technique as in \cite{Walgate02}.  For merits that will be revealed soon, we also show here that these states are PPT-indistinguishable.

\begin{figure}[t]
\centering
\includegraphics[scale=0.35]{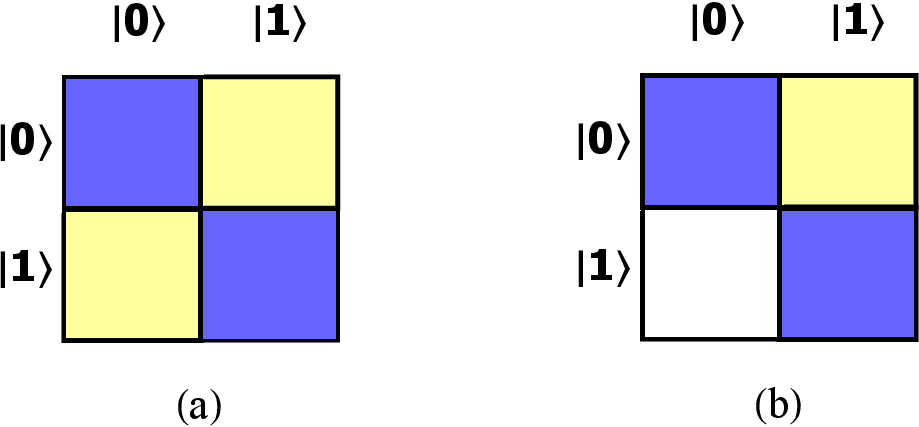}
\captionsetup{justification=raggedright}
\caption
{\small (a) The three Bell states (\ref{Bell}) in $\mathbb{C}^2 \otimes \mathbb{C}^2$; \ (b) States $\{|\beta_1\rangle, |\beta_2\rangle, |\gamma_3\rangle\}$ in $\mathbb{C}^2 \otimes \mathbb{C}^2$. \ The blue tiles signify $|\beta_{1, 2}\rangle$ and the yellow ones signify $|\beta_3\rangle$ or $|\gamma_3\rangle$.
}\label{fig1}
\end{figure}

\begin{Prop}\label{prop1}
The three orthogonal states $\{|\beta_1\rangle, |\beta_2\rangle,$ $|\gamma_3\rangle\}$ in $\mathbb{C}^2 \otimes \mathbb{C}^2$ are PPT-indistinguishable.
\end{Prop}

\begin{proof}
Assume contrarily that $\{|\beta_{1}\rangle, |\beta_{2}\rangle, |\gamma_{3}\rangle\}$ are PPT-distinguishable. Then by definition, there must exist PPT-POVMs $\{M_1, M_2, M_{3}\}$ (which means $M_k,$ $M_k^{\mathrm{T_A}} \geq 0$ for $1 \leq k \leq 3$) such that
\begin{equation}\label{Mi}
M_1 + M_2 + M_{3} = \mathbb{I}_{2 \otimes 2},
\end{equation}
and
\begin{equation*}\begin{split}
& \langle\beta_1|M_1|\beta_1\rangle = 1, \ \ \langle\beta_2|M_2|\beta_2\rangle = 1, \ \ \langle\gamma_3|M_3|\gamma_3\rangle = 1.
\end{split}\end{equation*}
By routine calculation, there is
\begin{equation*}\begin{split}
  & |\beta_{1,2}\rangle\langle\beta_{1,2}|^{\mathrm{T_A}}\\
& \ = \ \frac{|00\rangle\langle00| + |11\rangle\langle11| \pm (|01\rangle\langle10| + |10\rangle\langle01|)}{2}\\
& \ \leq \ \frac{|00\rangle\langle00| + |11\rangle\langle11| + (|01\rangle\langle01| + |10\rangle\langle10|)}{2} \ = \ \frac{\mathbb{I}}{2}.\\
\end{split}\end{equation*}
Hence, we get
\begin{equation*}\begin{split}
1 & = \mathrm{Tr}(M_{1,2} |\beta_{1,2}\rangle\langle\beta_{1,2}|) \\
  & = \mathrm{Tr}\left(M_{1,2}^{\mathrm{T_A}} |\beta_{1,2}\rangle\langle\beta_{1,2}|^{\mathrm{T_A}}\right) \leq \frac{1}{2} \ \mathrm{Tr}\left(M_{1,2}^{\mathrm{T_A}}\right)
\end{split}\end{equation*}
and $\mathrm{Tr}(M_{1,2}) = \mathrm{Tr}\left(M_{1,2}^{\mathrm{T_A}}\right) \geq 2$. However, by (\ref{Mi}) we have
\begin{equation*}
\mathrm{Tr}(M_1) + \mathrm{Tr}(M_2) + \mathrm{Tr}(M_3) = 4,
\end{equation*}
which forces $\mathrm{Tr}(M_3) = 0$. That is, $M_3 = 0$ and we get a contradiction. Therefore, the states $\{|\beta_{1}\rangle, |\beta_{2}\rangle, |\gamma_{3}\rangle\}$ are PPT-indistinguishable.
\end{proof}

In fact, one can easily observe from the proof that an arbitrary third orthogonal state $|\gamma'_3\rangle \in \mathbb{C}^2 \otimes \mathbb{C}^2$ suffices to make $\{|\beta_{1}\rangle, |\beta_{2}\rangle, |\gamma'_{3}\rangle\}$ PPT-indistinguishable. The advantage of proving PPT-indistinguishability of such states is that we can further extend the proof to higher dimensional situations where the third state is not entirely supported on $\mathbb{C}^2 \otimes \mathbb{C}^2$. Such extension seems improbable for methods like TOPLM.

\begin{Lem}\label{lem1}
In two-partite system $\mathbb{C}^{m} \otimes \mathbb{C}^{n} \supsetneq \mathbb{C}^2 \otimes \mathbb{C}^2$ (the $2\otimes 2$ product subspace on which $|\beta_{1, 2}\rangle$ is supported), the set of three orthogonal states $\{|\beta_{1}\rangle, |\beta_{2}\rangle, |\delta_{3}\rangle\}$ is  locally indistinguishable if and only if $|\delta_{3}\rangle$ has nonzero overlap with the subspace $\mathbb{C}^2 \otimes \mathbb{C}^2$:
\begin{equation}
\left(\sum_{i,j \in \{0,1\}}|ij\rangle\langle ij| \right) |\delta_{3}\rangle \neq 0.
\end{equation}
\end{Lem}

\begin{proof}
The ``only if'' direction is much obvious: if $\left(\sum_{i,j \in \{0,1\}}|ij\rangle\langle ij| \right) |\delta_{3}\rangle = 0$, then the two subsystems can make local projective measurements $\{|+\rangle\langle+|, ~|-\rangle\langle-|, ~\mathbb{I}_m - |+\rangle\langle+| - |-\rangle\langle-|\}$ and $\{|+\rangle\langle+|, ~|-\rangle\langle-|, ~\mathbb{I}_n-|+\rangle\langle+|-|-\rangle\langle-|\}$ to distinguish $\{|\beta_{1}\rangle, |\beta_{2}\rangle, |\delta_{3}\rangle\}$. For the ``if'' direction, we prove that $\{|\beta_{1}\rangle, |\beta_{2}\rangle, |\delta_{3}\rangle\}$ is PPT-indistinguishable. Suppose contrarily that these states are PPT-distinguishable. Then there must exist PPT-POVMs $\{M_1, M_2, M_3\}$ such that
\begin{equation}
M_1 + M_2 + M_3 = \mathbb{I}_{m \otimes n},
\end{equation}
and
\begin{equation*}\begin{split}
& \langle\beta_1|M_1|\beta_1\rangle = 1, \ \ \langle\beta_2|M_2|\beta_2\rangle = 1, \ \ \langle\delta_3|M_3|\delta_3\rangle = 1.
\end{split}\end{equation*}
Since $|\beta_{1, 2}\rangle$ are supported on the subspace $\mathbb{C}^2 \otimes \mathbb{C}^2$, we have $P|\beta_{1,2}\rangle = |\beta_{1,2}\rangle$ where $P=\left(\sum_{i,j \in \{0,1\}}|ij\rangle\langle ij| \right)$ is the projector onto the subspace $\mathbb{C}^2 \otimes \mathbb{C}^2$. Denote further that $\widetilde{M_i} = P M_i P$, then $\langle\beta_{1, 2}|\widetilde{M}_{1, 2}|\beta_{1, 2}\rangle = 1$. By the proof of Proposition \ref{prop1}, one can get $\mathrm{Tr}(\widetilde{M}_{1,2}) \geq 2$ and since
\begin{equation*}
\mathrm{Tr}(\widetilde{M_1}) + \mathrm{Tr}(\widetilde{M_2}) + \mathrm{Tr}(\widetilde{M_3}) = 4,
\end{equation*}
there must be $\widetilde{M}_3 = 0$. This further means $PM_3 = M_3P = 0$, which is obvious when we decompose $M_3$ as diagonal form. Now we get $M_3 = P^{\bot} M_3 P^{\bot}$ where $P^{\bot} = \mathbb{I}_{m\otimes n}-P$, but this leads to a contradiction:
\begin{equation*}\begin{split}
1 = \langle\delta_3|M_3|\delta_3\rangle & = \langle\delta_3|P^{\bot}M_3 P^{\bot}|\delta_3\rangle \\
& \leq \ \langle\delta_3|P^{\bot}|\delta_3\rangle = 1- \langle\delta_3|P|\delta_3\rangle < 1.
\end{split}\end{equation*}
As a consequence, the three states must be PPT-indistinguishable.
\end{proof}

With Lemma \ref{lem1}, we can now construct genuinely nonlocal sets of three in multi-qubit systems. We first show two examples and then a more general theorem for all $N$-qubit systems.

\begin{figure}[t]
\centering
\includegraphics[scale=0.35]{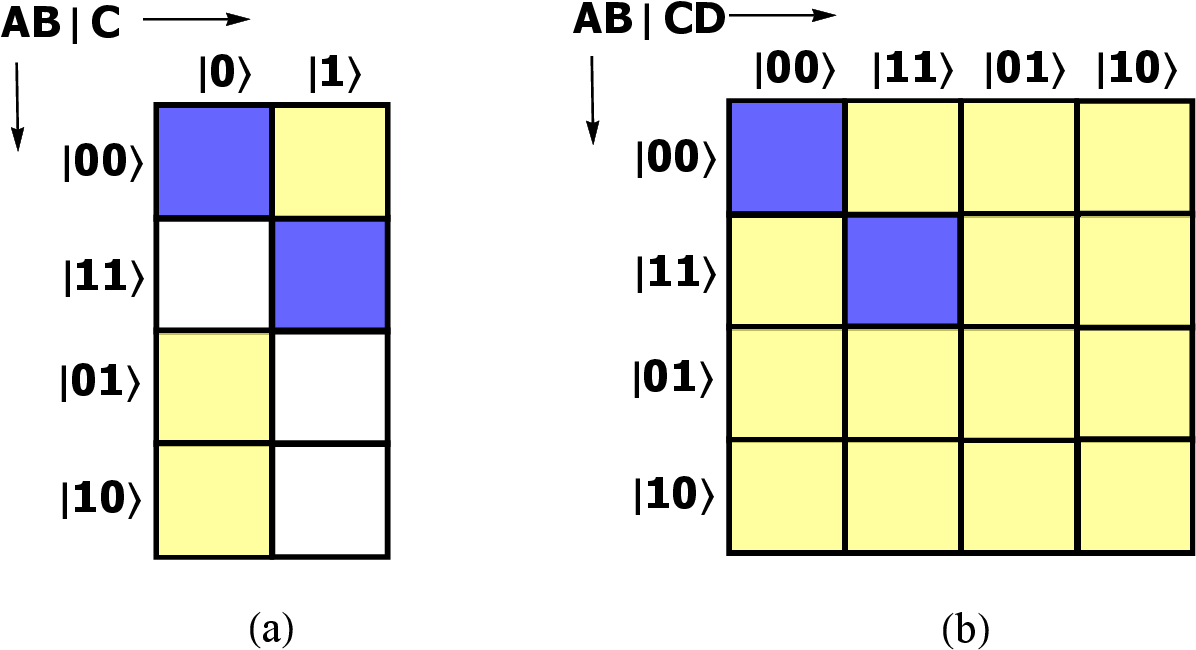}
\captionsetup{justification=raggedright}
\caption
{\small \ \ (a) States $\{|\varphi_1\rangle, |\varphi_2\rangle, |\varphi_3\rangle\}$ in bipartition AB$|$C; (b) States $\{|\phi_1\rangle, |\phi_2\rangle, |\phi_3\rangle\}$ in bipartition AB$|$CD. The blue tiles signify the GHZ pairs while the yellow ones signify the third states.
}\label{fig2}
\end{figure}

\begin{Exm}\label{exm1}
In $(\mathbb{C}^2)^{\otimes 3}$, the three orthogonal states consisting of the three-qubit GHZ pair
$$|\varphi_{1,2}\rangle = \frac{1}{\sqrt{2}} ( |000\rangle \pm |111\rangle )$$
and the W state
$$|\varphi_{3}\rangle = \frac{1}{\sqrt{3}} ( |001\rangle + |010\rangle + |100\rangle )$$
is genuinely nonlocal. Here, the third state $|\varphi_3\rangle$ has nonzero overlap with the $2 \otimes 2$ product subspace span$\{|00\rangle, |11\rangle\}$ $\otimes$ span$\{|0\rangle, |1\rangle\}$ in each of the bipartitions AB$|$C, AC$|$B and BC$|$A and thus Lemma \ref{lem1} can be applied. For example, see FIG \ref{fig2} (a) for bipartition AB$|$C.
\end{Exm}

\begin{Exm}\label{exm2}
In $(\mathbb{C}^2)^{\otimes 4}$, the three orthogonal states consisting of the four-qubit GHZ pair
$$|\phi_{1,2}\rangle = \frac{1}{\sqrt{2}} ( |0000\rangle \pm |1111\rangle ) $$
and the third orthogonal state
\begin{equation*}\begin{split}
|\phi_{3}\rangle = \frac{1}{\sqrt{14}} & \left( |0001\rangle + |0010\rangle + |0011\rangle + \cdots \phantom{a^{b}\cdots} \right.\\
& \left. \phantom{a^{b}\cdots} \cdots + |1100\rangle + |1101\rangle + |1110\rangle\right)
\end{split}\end{equation*}
is also genuinely nonlocal. Similarly, the third state $|\phi_3\rangle$ has nonzero overlap with the $2 \otimes 2$ product subspaces span$\{|000\rangle, |111\rangle\}$ $\otimes$ span$\{|0\rangle, |1\rangle\}$ and span$\{|00\rangle, |11\rangle\}$ $\otimes$ span$\{|00\rangle, |11\rangle\}$, across the four ``3--1'' bipartitions and the three ``2--2'' bipartitions respectively. And then, Lemma \ref{lem1} can be applied. For example, see FIG \ref{fig2} (b) for the indistinguishability within bipartition AB$|$CD.
\end{Exm}

\begin{Thm}\label{thm1}
In $N$-qubit system $(\mathbb{C}^2)^{\otimes N}$, the set consisting of the $N$-qubit GHZ pair
\begin{equation*}\begin{split}
|\Gamma_{1,2}\rangle = \frac{1}{\sqrt{2}} \left( |0\rangle^{\otimes N} \pm |1\rangle^{\otimes N} \right) \\
\end{split}\end{equation*}
and a third state
\begin{equation*}
|\Gamma_{3}\rangle =\sum_{i_1 i_2 \cdots i_N\in \mathcal{P}} c_{i_1 i_2 \cdots i_N} |i_1 i_2 \cdots i_N\rangle
\end{equation*}
is genuinely nonlocal, where the index set $\mathcal{P} = \{0,1\}^N \backslash \{00\cdots 0, 11\cdots 1\}$ and the coefficients $c_{i_1 i_2 \cdots i_N}$ of $|\Gamma_{3}\rangle$ satisfy:
\begin{equation}\label{overlap}
|c_{i_1 i_2 \cdots i_N}|^2 + |c_{_{\overline{i_1 i_2 \cdots i_N}}}|^2  > 0
\end{equation}
and
\begin{equation}
\sum_{i_1 i_2 \cdots i_N \in \mathcal{P}} |c_{i_1 i_2 \cdots i_N}|^2 = 1.
\end{equation}
\end{Thm}

\medskip

\begin{proof}
For any $i_1 i_2 \cdots i_N \in \mathcal{P}$, we denote its bit-wise reverse as $\overline{i_1 i_2 \cdots i_N}$. Then each pair of $i_1 i_2 \cdots i_N $ and $\overline{i_1 i_2 \cdots i_N }$ corresponds to one of the $|\mathcal{P}|/2 = 2^{N-1}-1$ bipartitions of the $N$ subsystems. Because of (\ref{overlap}), the third state $|\Gamma_{3}\rangle$ must have nonzero overlap with a subspace $\mathbb{C}^2 \otimes \mathbb{C}^2$ (on which $|\Gamma_{1,2}\rangle$ is supported) in each of the bipartitions. Therefore, genuine nonlocality of the three states $\{|\Gamma_1\rangle, |\Gamma_2\rangle, |\Gamma_3\rangle\}$ can be deduced from Lemma \ref{lem1}.
\end{proof}

Note that in the above construction, the third state is not necessary a genuinely entangled state: taking $N = 3$ for exmaple, the third state can be constructed as $|\varphi'_{3}\rangle = |0\rangle \otimes \frac{1}{\sqrt{3}} ( |11\rangle + |10\rangle + |01\rangle)$. Nevertheless, the most natural choice for the coefficients $c_{i_1 i_2 \cdots i_N}$ produce genuinely entangled ones.

\begin{Cor}
In $N$-qubit system $(\mathbb{C}^2)^{\otimes N}$, the following three orthogonal genuinely entangled states are genuinely nonlocal:
\begin{equation}\label{ge3}
\begin{split}
&|\Gamma_{1,2}\rangle = \frac{1}{\sqrt{2}} \left( |0\rangle^{\otimes N} \pm |1\rangle^{\otimes N} \right) \\
\\
&|\Lambda_{3}\rangle = \frac{1}{\sqrt{|\mathcal{P}|}} \sum_{i_1 i_2 \cdots i_N \in \mathcal{P}} |i_1 i_2 \cdots i_N \rangle.\\
\end{split}
\end{equation}
\end{Cor}

\begin{proof}
By Theorem \ref{thm1}, these states are genuinely nonlcal, so it remains to show that $|\Lambda_{3}\rangle$ is genuinely entangled. Without loss of generality, assume that $|\Lambda_{3}\rangle = |\theta\rangle \otimes |\xi\rangle$ in bipartition $\{1, \cdots, s\} \ | \ \{s+1, \cdots, N\}$, where $$|\theta\rangle = \sum_{i_1 \cdots i_s \in \{0, 1\}^s} a_{i_1 \cdots i_s} \ |i_1 \cdots i_s\rangle$$ and $$|\xi\rangle = \sum_{j_1 \cdots j_{N-s} \in \{0, 1\}^{N-s}} b_{j_1 \cdots j_{N-s}} \ |j_1 \cdots j_{N-s}\rangle.$$ By the construction of $|\Lambda_{3}\rangle$, we have $a_{\overrightarrow{0}} \cdot b_{\overrightarrow{0}} =0$ and $a_{\overrightarrow{1}} \cdot b_{\overrightarrow{1}} =0$, where $\overrightarrow{0}$, $\overrightarrow{1}$ are vectors of all $0$ and all $1$. But we also have $a_{\overrightarrow{0}} \cdot b_{\overrightarrow{1}} = a_{\overrightarrow{1}} \cdot b_{\overrightarrow{0}} = 1/\sqrt{|\mathcal{P}|}$, which leads to a contradiction. Therefore, $|\Lambda_{3}\rangle$ cannot be a product state in any bipartition and it is genuinely entangled.
\end{proof}

Notice that $(\mathbb{C}^2)^{\otimes N}$ is the smallest $N$-partite systems, which can be embedded into any $N$-partite system with larger local dimension. Therefore, genuinely nonlocal sets of three also exist in arbitrary $N$-partite systems.

For a second corollary of Theorem \ref{thm1}, it turns out that the three states in (\ref{ge3}) are also strongly nonlocal, owing to the following proposition that was established by Halder et al. in their original work about ``strong nonlocality without entanglement''.

\begin{Prop}[\cite{Halder19}]
Any set of three locally indistinguishable orthogonal pure states must be locally irreducible.
\end{Prop}

In that work, the authors also questioned about the the existence of strong nonlocality with genuine entanglement. The works \cite{Wang21, Hu24} already answered this positively, yet their constructions are quite complecated and only specialized for some systems. Some other works also dived into the constructions of strongly nonlocal sets with smaller cardinality \cite{liji,Hu24,Zhen24} and the problem whether strongly nonlocal sets exist in all multipartite systems \cite{Shi22, Zhou23, He24}. With no sweat, our construction in (\ref{ge3}) settles these problems all together:

\begin{Cor}
In any $N$-partite system with arbitrary local dimensions, there always exist three orthogonal genuinely entangled states which are strongly nonlocal.
\end{Cor}

Such a result is somewhat unexpected, in the context that all existing examples of strongly nonlocal sets found by researchers have cardinality dramatically larger, as all of them are raised with TOPLM \cite{Li23}. This  potentially illuminates future exploration for new types of strongly nonlocal sets utilizing techniques other than TOPLM.

\section{Mixed hypothetical states: existence of genuinely nonlocal sets of two, even in the many copy limit}\label{section4}
In Section \ref{section3}, a central feature of genuine nonlocality of the multipartite pure states is that the subsystem observers share only one copy of the unknown state. In fact, it is only under the ``single copy'' constraint that the nonlocal nature of quantum information is manifested, for it was proved by Bandyopadhyay in \cite{Bandyopadhyay11} that any set of orthogonal pure states can be locally distinguished if sufficiently many copies of the unknown state are available. Nevertheless, it was also discovered in the same paper that orthogonal mixed states can still manifest local indistinguishability even in the many copy limit. Taking advantage of certain special properties of the unextendible product bases (UPBs) \cite{Bennett1999,DiVicenzo03}, Bandyopadhyay constructed examples of bipartite orthogonal states, not all of which are pure, that cannot be distinguished locally even if $n$ copies of the unknown state are available, where $n$ is finite but arbitrarily large. Remarkably, such sets can even be minimal in cardinality, consisting of only two orthogonal states.

Back to multipartite systems, although pure states fail to exhibit genuine nonlocality in the many copy limit, it's natural to conjecture that orthogonal mixed states can behave differently, just as the bipartite scenario. However, one would immediately doubt whether the minimal cardinality two is achievable, as genuine nonlocality seems much stronger than local indistinguishability. In what follows, we are to provide examples that not only confirm the former speculation but also answer the latter question positively. Namely, for $N$-partite systmes with arbitrary $N$, there always exist two orthogonal mixed states that cannot be distinguished by the subsystems unless all of them join together, regardless of how many copies of the state are available. Our construction primarily follows the same vein as \cite{Bandyopadhyay11}, but it's noteworthy that we can provide examples for multipartite systems with arbitrary local dimensions, while Bandyopadhyay's original construction for the bipartite case only works for $\mathbb{C}^m \otimes \mathbb{C}^n$ with $m, n \geq 3$ as it is based on UPBs, which do not exist in $\mathbb{C}^2 \otimes \mathbb{C}^d$ \cite{DiVicenzo03}.

Two propositions are crucial to the construction of Bandyopadhyay's and also ours:
\begin{Prop}[\cite{DiVicenzo03}]\label{prop3}
Let $S_1 = \{|\varphi_i\rangle\}_{i=1}^{k_1}$ and $S_2 = \{|\psi_j\rangle\}_{j=1}^{k_2}$ be two sets of bipartite UPB on $\mathcal{H}_{A_1} \otimes \mathcal{H}_{B_1}$ and $\mathcal{H}_{A_2} \otimes \mathcal{H}_{B_2}$, then $S_1 \otimes S_2 \triangleq \{|\varphi_i\rangle \otimes |\psi_j\rangle\}_{i, j = 1}^{k_1, k_2}$ is a bipartite UPB on $\mathcal{H}_{A_1A_2} \otimes \mathcal{H}_{B_1B_2}$.
\end{Prop}

\begin{Prop}[\cite{Chefles04}]\label{prop4}
If a set of orthogonal quantum states $\{\rho_1, \cdots, \rho_k\}$ in any bipartite system $ \mathcal{H}_A \otimes \mathcal{H}_B$ is conclusively locally distinguishable by LOCC, then for each $i \in \{1, \cdots, k\}$ there exist a product state $|\phi_i\rangle \in \mathcal{H}_A \otimes \mathcal{H}_B$ such that $\langle \phi_i|\rho_j |\phi_i \rangle = 0$ $\forall j \neq i$ and $\langle \phi_i|\rho_i |\phi_i \rangle > 0$ \cite{bipartite}.
\end{Prop}

In Bandyopadhyay's construction, a set of UPB $S= \{|\eta_i\rangle\}_{i=1}^{k}$ on bipatite system $\mathcal{H}_{A} \otimes \mathcal{H}_{B}$ is first specified. Then, the first state is constructed as $\sigma = 1/k \sum_{i=1}^k |\eta_i \rangle\langle \eta_i|$, which is also the normalized projector onto the subspace $\mathcal{H}_S = \text{span}\{|\eta_i\rangle\}_{i=1}^{k} \varsubsetneq \mathcal{H}_{A} \otimes \mathcal{H}_{B}$ and the second state $\rho$ can be any one supported on $\mathcal{H}_S^{\perp}$, the orthogonal complement of $\mathcal{H}_S$. To distinguish $\sigma$ and $\rho$ provided $n$ copies, one can of course make independent measurements on each copy (adaptively or non-adaptively), but those are special cases of distinguishing $\sigma^{\otimes n}$ and $\rho^{\otimes n}$ with the more general collective measurements. Note that $\sigma^{\otimes n}$ is the projector onto the subspace $\mathcal{H}_S^{\otimes n} = \text{span } S^{\otimes n}$, with $S^{\otimes n}$ being a bipartite UPB on $\mathcal{H}_{A}^{\otimes n} \otimes \mathcal{H}_{B}^{\otimes n}$ due to Proposition \ref{prop3}. This implies the nonexistence of any product state $|\phi\rangle$ satisfying $\langle \phi|\sigma^{\otimes n} |\phi \rangle = 0$ and therefore $\rho^{\otimes n}$ cannot be conclusively distinguished from $\sigma^{\otimes n}$ because of Proposition \ref{prop4}.

Illuminated by the bipartite scenario, in multipartite system $\mathcal{H}_1 \otimes \mathcal{H}_2 \otimes \cdots \otimes \mathcal{H}_N$, \ if there is some set of orthogonal fully product states $\{|\zeta_i\rangle\}_{i=1}^k$ that is unextendible in any bipartition of the subsystems, then the projector $\sigma = 1/k \sum_{i=1}^k |\zeta_i \rangle\langle \zeta_i|$ and another orthogonal state $\rho$ must be also locally indistinguishable in all biparitions. In fact, multipartite product states with this property are called genuinely unextendible product bases (GUPBs). However, by now there are only some negative results about their existence \cite{Demianowicz22,Shi23} and not even one example has ever been found by researchers. Fortunately, it was pointed out in \cite{Demianowicz18} that the orthogonality condition is unnecessary for us to construct subspaces where fully product states or biproduct states are absent.

A set of fully product states $S = \{|\Psi_j\rangle = |\alpha_j^1\rangle \otimes |\alpha_j^2\rangle \otimes \cdots \otimes |\alpha_j^N\rangle\}_{j=1}^k$ in $\mathcal{H} = \bigotimes_{i=1}^N \mathcal{H}_i$, not necessarily orthogonal, is called an nonorthogonal unextendible product basis (nUPB) if it spans a proper subspace $\mathcal{H}_S \varsubsetneq \mathcal{H}$ with no fully product state exists in its orthogonal complement $\mathcal{H}_S^{\perp}$. To construct such sets $S$ with the additional property of unextendibility in every bipartition (we called them nGUPBs in what follows), one may leverage the following ``local spanning'' condition that has been employed frequently in the literature for constructing orthogonal UPBs:

\begin{Prop}[\cite{DiVicenzo03}]\label{prop5}
For a set of $k \geq m+n-1$ product states $S = \{|\varphi_j\rangle \otimes |\psi_j\rangle\}_{j=1}^k$ in system $\mathbb{C}^m \otimes \mathbb{C}^n$, if every $m$ among $\{|\varphi_j\rangle\}_{j=1}^k$ span $\mathbb{C}^m$ and every $n$ among $\{|\psi_j\rangle\}_{j=1}^k$ span $\mathbb{C}^n$, then there must not exist any product state orthogonal to those in $S$.
\end{Prop}

The reason behind is simple: if a product state $|\theta\rangle \otimes |\xi\rangle$ is orthogonal to $S$, then by the local spanning condition $|\theta\rangle$ can at most be orthogonal to $m-1$ among $\{|\varphi_j\rangle\}_{j=1}^k$ and hence $|\xi\rangle$ must be orthogonal to the other $k-(m-1) \geq n$ among $\{|\psi_j\rangle\}_{j=1}^k$, which contradicts the local spanning condition on $\mathbb{C}^n$. Taking advantage of this, \cite{Demianowicz2022} proposed a succinct and general construction of nGUPBs in arbitrary multipartite system $\mathcal{H} = \bigotimes_{i=1}^N \mathbb{C}^{d_i}$ based on the Vandermonde matrix
\begin{equation*}
V_{k, D} = \arraycolsep = 0.33em \left(\begin{array}{c} \begin{matrix}
1 & x_1 & \cdots & x_1^{D-1} \\
1 & x_2 & \cdots & x_2^{D-1} \\
\vdots & \vdots & \ddots & \vdots \\
1 & x_{k} & \cdots & x_{k}^{D-1} \\
\end{matrix}\end{array}\right)
\end{equation*}
where $D = \prod_{i=1}^N d_i$ is the global dimension and $k \geq D/d_{min} + d_{min} - 1$. The variables $x_1, x_2, \cdots, x_k$ can be evaluated as $x_j = \exp(j\cdot\frac{2\pi i}{p})$ with some prime $p > D$ which ensure that $V_{k, D}$ is totally nonsingular, namely all its square submatrices are nonsingular. The nGUPB is constructed as:
\begin{equation*}\begin{split}
 & |\Psi_j\rangle = \frac{1}{\sqrt{D}}\sum_{l=0}^{D-1} x_j^{l} \ |l\rangle  \\
 = &\ \frac{1}{\sqrt{D}}\sum_{l_1 = 0}^{d_1-1} \sum_{l_2 = 0}^{d_2-1} \cdots \sum_{l_N = 0}^{d_N-1} x_j^{l_1 \frac{D}{d_1} + l_2 \frac{D}{d_1d_2} + \cdots + l_N} |l_1 l_2 \cdots l_N\rangle \\
 = &\ \bigotimes_{i=1}^{N} \ \frac{1}{\sqrt{d_i}}\sum_{l_i = 0}^{d_i-1} \left(x_j^{d_{i+1}\cdots d_{N}}\right)^{l_i} |l_i\rangle \ \ \ \ \ (j = 1, 2, \cdots, k).
\end{split}\end{equation*}
Such states satisfy the local spanning condition in every bipartition for if $|\Psi_j\rangle = |\varphi_j\rangle_L \otimes |\psi_j\rangle_{R}$ $(j = 1, 2, \cdots, k)$, then both
$$\frac{1}{\sqrt{\prod_{i\in L}d_n}}\sum_{j=1}^k |\varphi_j\rangle_L \langle j|$$
and
$$\frac{1}{\sqrt{\prod_{i\in R}d_n}}\sum_{j=1}^k |\psi_j\rangle_{R} \langle j|$$
are submatrices of $V_{k, D}^T$ and hence they are also totally nonsigular. This means any $\prod_{i\in L}d_i$ among $\{|\varphi_j\rangle_L\}_{j=1}^k$ span $\bigotimes_{i\in L}\mathbb{C}^{d_i}$ and likewise for $\{|\psi_j\rangle_R\}_{j=1}^k$.

Based on the above, the set of $N$-partite mixed states consisting of
\begin{equation}
\sigma = \frac1k \sum_{j=1}^k|\Psi_j\rangle\langle\Psi_j|
\end{equation}
(or any other state full-rank supported on $\mathcal{H}_S = \text{span}\{|\Psi_j\rangle\}_{j=1}^k$) and a second state
\begin{equation}
\rho \in \mathcal{H}_S^{\perp}
\end{equation}
is genuinely nonlocal. To complete our statement, it remains to show that genuine nonlocality persists in the many copy limit. For this, we need the following nonorthogonal version of Proposition \ref{prop3}:

\begin{Lem}\label{lem2}
If $S_1 = \{|\varphi_i\rangle\}_{i=1}^{k_1}$ and $S_2 = \{|\psi_j\rangle\}_{j=1}^{k_2}$ are two sets of nUPB on $\mathcal{H}_{A_1} \otimes \mathcal{H}_{B_1}$ and $\mathcal{H}_{A_2} \otimes \mathcal{H}_{B_2}$, then $S_1 \otimes S_2$ is a nUPB on $\mathcal{H}_{A_1A_2} \otimes \mathcal{H}_{B_1B_2}$.
\end{Lem}

The proof is mostly identical to that of Proposition \ref{prop3} and one may refer to \cite{DiVicenzo03}. For arbitrarily large $n$, since $\sigma^{\otimes n}$ is full-rank supported on the subspace $\mathcal{H}_S^{\otimes n}$ spanned by the nUPB $\left(\{|\Psi_j\rangle\}_{j=1}^k\right)^{\otimes n}$, in every bipartition it cannot be conclusively distinguished from $\rho^{\otimes n}$ by Proposition \ref{prop4}. That is to say:
\begin{Thm}\label{thm2}
In any $N$-partite system with arbitrary local dimensions, there always exist two orthogonal mixed states which are genuinely nonlocal in the many copy limit.
\end{Thm}

Notice that in our construction, the second state $\rho$ is necessarily genuinely entangled, for biproduct states are absent in the subspace $\mathcal{H}_S^{\perp}$. On the other hand, although $\sigma$ is fully separable (meaning that it does not possess any entanglement in any bipartiton), such states are actually not necessary. For example, the first state can also be constructed as $\sigma' = (1-\varepsilon) \sigma + \varepsilon|\chi\rangle \langle\chi|$, where $|\chi\rangle \in \mathcal{H}_S^{\perp}$ and the second state can be any one orthogonal to $\sigma'$, in which case both states are genuinely entangled.

\section{Conclusions}\label{section5}
In this work, we consider the problem of constructing genuinely nonlocal sets of orthogonal $N$-partite quantum states with the smallest possible cardinality. For pure hypothetical states, we show that genuinely nonlocal sets of three exist in all $N$-partite systems. This also gives the first examples of strongly nonlocal sets that are not constructed through the TOPLM technique and have dramatically smaller cardinality then ever. For mixed hypothetical states, we show that genuinely nonlocal sets of two exist for arbitrary $N$-partite system. Moreover, upon such states genuine nonlocality persists even in the many copy limit. Noteworthily, since certain genuinely multipartite entangled states are necessary in our achievement to such minimum cardinality, it potentially indicates some nontrivial ability of genuine multipartite entanglement for raising difficulty in locally accessing multipartite quantum information.

In technical aspect, our results also indicate some evident limitation of the TOPLM technique for detecting multipartite nonlocality: in $(\mathbb{C}^d)^{\otimes N}$ TOPLM can only detect genuinely (strongly) nonlocal sets with cardinality no smaller than $d^{N-1}+1$, while here we show that genuinely (strongly) nonlocal sets with cardinality down to three exist. This might illuminate future research directions for developing new techniques other than TOPLM for more effective detection of such nonlocality.

There are also some questions left to be considered. For example, is that possible for two low-rank mixed states to manifest genuine nonlocality? Note that the rank of the first state $\sigma$ in our construction in Section \ref{section4} is at least $D/d_{min} + d_{min} - 1$, while on the other end genuine nonlocality can never be manifested upon two pure states (which are rank-one). Besides, it's natural to question the existence of strongly nonlocal sets of four, five, $\cdots$ in $N$-partite systems, as the supersets of a locally irreducible set may not be locally irreducible. Also, it would be interesting to ask whether genuinely nonlocal sets with smallest cardinality can be achieved without the presence of genuine multipartite entanglement. Though there seems to be a slight possibility when $N$ is small (say $N=3$), we conjecture that it is impossible when $N$ is large.

\medskip
\centerline{\textbf{ACKNOWLEDGMENTS}}
The authors of this work were supported by Quantum Science and Technology - National Science and Technology Major Project under Grant No. 2024ZD0300500, the National Natural Science Foundation of China (Grant No.62401638, No.12447107, No.12371458, No.92465202, No.62272492), the Guangdong Provincial Quantum Science Strategic Initiative under Grant No.GDZX2303007 and the Guangzhou Science and Technology Program under Grant No.2024A04J4892.


\begin{thebibliography}{}
\bibitem{bipartite}
The original proposition works for multipartite systems but here only the bipartite version is needed.
\end{thebibliography}
\end{document}